\documentclass[twocolumn,superscriptaddress,amsmath,amssymb,showpacs,prl]{revtex4-1}

\usepackage{graphicx}

\usepackage{color}
\renewcommand{\phi}{\varphi}

\begin{document}

\title{Unveiling the effect of Ni on the formation and structure of Earth's inner core }

\author{Yang Sun}
    \email{yangsun@xmu.edu.cn}
    \affiliation{Department of Physics, Xiamen University, Xiamen, Fujian 361005, China}
	\affiliation{Department of Applied Physics and Applied Mathematics, Columbia University, New York, NY 10027, USA}
    \affiliation{Department of Physics, Iowa State University, Ames, IA 50011, USA}
    
\author{Mikhail I. Mendelev}
    \email{mikhail.mendelev@gmail.com}
	\affiliation{Department of Physics, Iowa State University, Ames, IA 50011, USA}
 
\author{Feng Zhang}
    \affiliation{Department of Physics, Iowa State University, Ames, IA 50011, USA}
	\affiliation{Ames Laboratory, US Department of Energy, Ames, IA 50011, USA}

\author{Xun Liu}
	\affiliation{Research and Services Division of Materials Data and Integrated System, National Institute for Materials Science, Ibaraki 305-0044, Japan}
\author{Bo Da}
	\affiliation{Research and Services Division of Materials Data and Integrated System, National Institute for Materials Science, Ibaraki 305-0044, Japan}

\author{Cai-Zhuang Wang}
	\affiliation{Ames Laboratory, US Department of Energy, Ames, IA 50011, USA}
    \affiliation{Department of Physics, Iowa State University, Ames, IA 50011, USA}
 
\author{Renata M. Wentzcovitch}
    \email{rmw2150@columbia.edu}
	\affiliation{Department of Applied Physics and Applied Mathematics, Columbia University, New York, NY 10027, USA}
	\affiliation{Department of Earth and Environmental Sciences, Columbia University, New York, NY 10027, USA}
	\affiliation{Lamont–Doherty Earth Observatory, Columbia University, Palisades, NY 10964, USA}
 
\author{Kai-Ming Ho}
	\affiliation{Department of Physics, Iowa State University, Ames, IA 50011, USA}

\date{Sep. 15, 2023}

\begin{abstract}

\textbf{Ni is the second most abundant element in the Earth's core. Yet, its effects on the inner core's structure and formation process are usually disregarded because of its electronic and size similarity with Fe. Using \emph{ab initio} molecular dynamics simulations, we find that the bcc phase can spontaneously crystallize in liquid Ni at temperatures above Fe's melting point at inner core pressures. The melting temperature of Ni is shown to be 700-800 K higher than that of Fe at 323-360 GPa. hcp, bcc, and liquid phase relation differ for Fe and Ni. Ni can be a bcc stabilizer for Fe at high temperatures and inner core pressures. A small amount of Ni can accelerate Fe's crystallization at core pressures. These results suggest Ni may substantially impact the structure and formation process of the solid inner core.}

\end{abstract}
\maketitle

The Earth has a liquid outer core and a solid inner core composed of Fe with a small amount of Ni and light elements. The solid phase's chemical composition and structure are fundamental for understanding the core, but there are still uncertainties \cite{1,2,3}. Fe alloys in the inner core are often believed to have the hexagonal close-packed (hcp) lattice\cite{4,5}, while the body-centered cubic (bcc) phase is also under consideration\cite{6,7}. Thermodynamic calculations indicate that hcp Fe is the stable solid phase at inner core pressures. Still, the Gibbs free energy difference between the hcp and bcc phases may be much smaller than previously thought \cite{8,9,10}. The presence of other elements diluted in Fe may significantly change the bcc and hcp stability fields \cite{11,12}. The growth of the present inner core provides the primary power source to sustain the outer core convection, which generates the Earth's magnetic field \cite{13}. Despite its importance, significant gaps exist in understanding the inner core's age and nucleation process \cite{14,15,16,17}. Recent simulations showed that the hcp Fe nucleation requires much larger supercooling than the liquid core can reach, leading to the "inner core nucleation paradox." \cite{18,19,20,21} Metastable bcc Fe was found to nucleate at smaller supercooling than hcp Fe at core pressures, which may help resolve the paradox \cite{21}. 

Cosmochemical models estimated 5-15 wt.$\%$ Ni content in the core \cite{1}. Iron meteorites can contain even $\sim$35 wt.$\%$ Ni \cite{22}. High-pressure experiments of Fe-Ni alloys focused mainly on the solid phase relations. Ni is known to stabilize the fcc phase with respect to hcp under low P-T conditions \cite{23,24,25,26}. The bcc phase was once reported in a Fe$_{90}$Ni$_{10}$ alloy by experiments at pressures above 225 GPa and temperatures over 3400 K \cite{27}. However, experiments with P-T conditions up to 340 GPa and 4700 K did not confirm the existence of the bcc phase in Fe$_{90}$Ni$_{10}$ \cite{28,29}. Static calculations demonstrated that Ni doping could improve the dynamic stability of bcc Fe \cite{30}. In recent experiments, Ni and FeNi melting curve was studied up to 120 GPa \cite{31,32,33}. It was suggested that Ni could strongly modify the hcp/fcc/liquid triple points in the Fe$_{1-x}$Ni$_x$ alloy \cite{33}. So far, the effect of Ni on the inner core's nucleation process is yet to be considered.

Previous \emph{ab initio} quasiharmonic calculations of Ni and Fe-Ni alloys at Earth's core pressures focused mainly on the fcc-hcp phase relation and assumed anharmonicity to be negligible\cite{19}. However, the inner core and outer core are under a solid-liquid coexistence condition with temperatures close to the melting point. Anharmonic effects can significantly affect the free energies and phase stabilities, especially in the bcc phase \cite{21,34,35}. One practical way to determine the stable phase near the melting point is to compute the melting temperatures of competing crystalline phases when the stable solid phase is uncertain. These melting temperatures define the relative stability of different solid phases for liquid coexistence: the phase with the highest melting temperature is the most stable, while other crystalline phases are metastable. In this work, we start with an unexpected simulation result, i.e., the crystallization of liquid Ni at inner core boundary (ICB) conditions. Then we address Ni's hcp, fcc, and bcc phase stability near the melting temperature and compare it with the corresponding data for Fe. Combining free energy results and crystallization kinetics of FeNi systems, we discuss possible scenarios for the effect of Ni on the formation and structure of the inner core.

\begin{figure}
\includegraphics[width=0.5\textwidth]{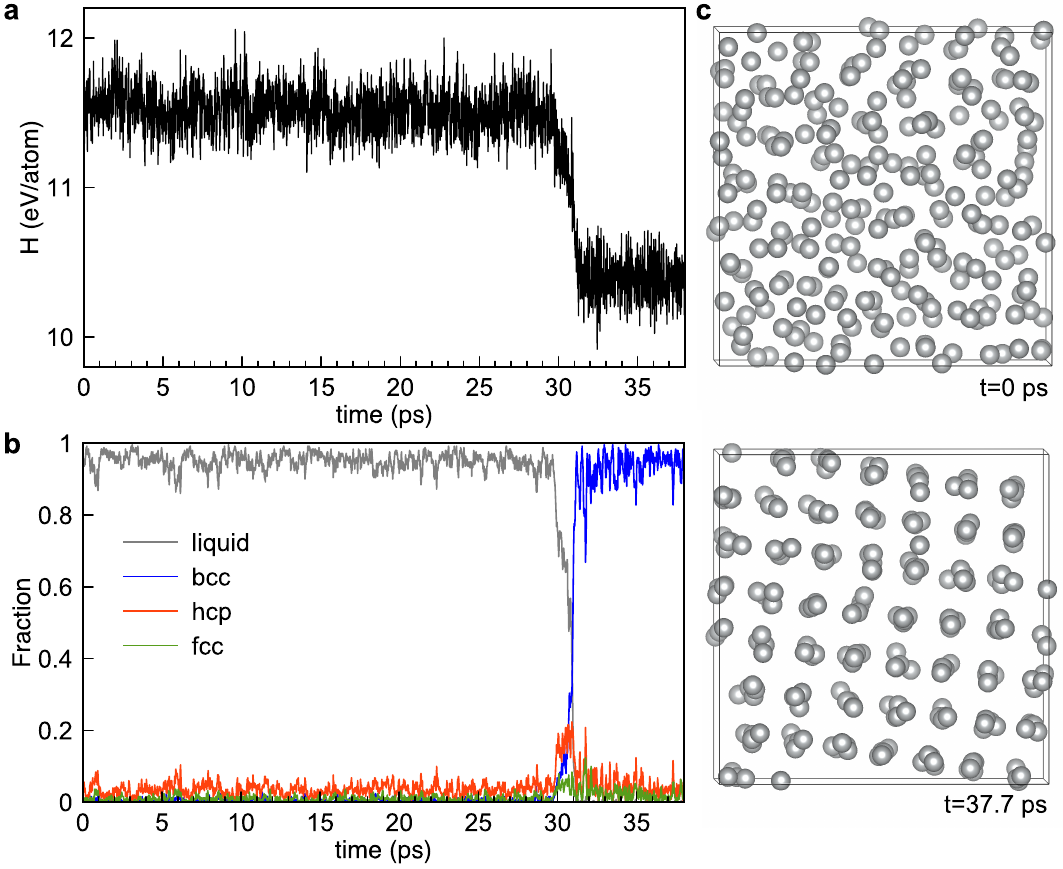}
\caption{\label{fig:fig1} Crystallization from liquid to bcc at 323 GPa and 6000 K by the AIMD simulation. a. Enthalpy as a function of simulation time. b. The fraction of liquid and crystalline atoms as a function of simulation time. c. Initial (upper) and final (bottom) atomic configurations with 250 atoms. }
\end{figure}

\textbf{Bcc crystallization.}
Our \emph{ab initio} calculations showed that the melting temperature of hcp Fe is 5848 K at 323 GPa \cite{10}, which is consistent to previous calculations of 5730±200 K with the same pseudopotential (see Methods) \cite{36}. Including more valence electrons in the pseudopotential systematically increases the melting points but does not significantly change the relative stability between the hcp and bcc phases  \cite{10}(see Methods). If Ni were similar to Fe, as usually assumed, one would expect a similar melting temperature for Ni at the same pressures. Surprisingly, we found that Ni liquid crystallizes spontaneously in an \emph{ab initio} molecular dynamics (AIMD) simulation at 6000 K and 323 GPa, which is well above Fe's melting point. Figure 1a shows the enthalpy change as a function of time during an AIMD simulation. The sharp drop of the enthalpy at $\sim$30 ps indicates a clear first-order phase transition. Figure~\ref{fig:fig1}b shows that the fraction of the bcc phase quickly increases at $\sim$30 ps, which coincides with the enthalpy change in Fig.~\ref{fig:fig1}a. The fraction of hcp also increases at 30 ps and stays steady to 31 ps but quickly decreases after the bcc phase becomes dominant. The fraction of fcc during the crystallization is insignificant. The phase competition is mainly between hcp and bcc during the nucleation process. The initial and final snapshots shown in Fig.~\ref{fig:fig1}c confirm the liquid solidified into the bcc phase during the AIMD simulation. The solidification of Ni at T=6000 K was unexpected because this temperature is well above Fe's melting temperature. Therefore, the Ni melting temperature should be significantly higher than Fe at 323 GPa. The crystallization of the bcc phase also indicates bcc has no instability at 6000 K, which is different from the previously revealed instability in the bcc phase at 0 K \cite{37}. Therefore, anharmonicity contributes significantly to the bcc phase stability at high temperatures.

Crystallization is a rare event challenging to observe spontaneously in simulations. Based on the classical nucleation theory \cite{38}, the nucleation rate depends exponentially on the nucleation barrier $\Delta G^*$, as $J= \kappa \text{exp} ( - \Delta G^{*} / k_B T ) $, where $\kappa$ is a kinetic prefactor. The fast crystallization observed here indicates a significant nucleation rate, thus a small nucleation barrier. The key factors that determine the nucleation barrier $\Delta G^{*}$ are the free energy difference between the bulk solid and liquid phases and the solid-liquid interface (SLI) free energy. While the bcc phase prevails over hcp in the crystallization process, it does not necessarily mean that bcc Ni has significantly lower free energy than hcp Ni. In the case of Fe, metastable bcc can have a nucleation rate 10-40 orders of magnitude higher than that of the stable hcp phase at 1000-600 K supercooling \cite{21}, because the SLI free energy for Fe's bcc phase is much smaller than that for the hcp phase \cite{21}. Therefore, one needs the  free energy relations to determine the relative thermodynamic stability of bcc, hcp, and liquid. As free energy relations also define melting temperatures, we use the latter to infer thermodynamic stability among these phases. 

\begin{figure}
\includegraphics[width=0.45\textwidth]{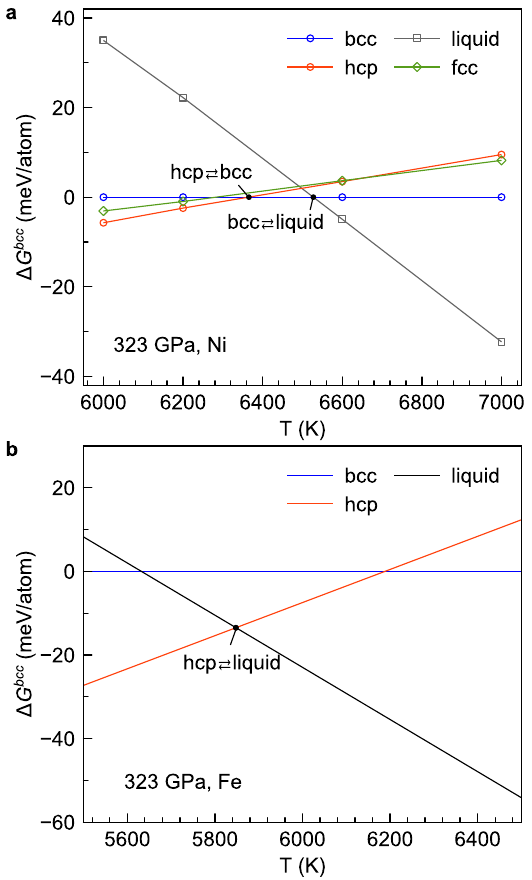}
\caption{\label{fig:fig2} \textbf{\emph{Ab initio} free energy difference referenced to bcc phase at 323 GPa near the melting temperatures for Ni and Fe.} a. The open circles are obtained through TI calculations. The circles are connected to guide the eyes. The two black dots at 6528 K and 6366 K indicate hcp-bcc and bcc-liquid phase transitions, respectively. b. The free energy data of Fe are from Ref. \cite{10}. The black dot indicates the phase transition of hcp-liquid. The uncertainty is $\sim$ 1 meV/atom due to the size effect \cite{10}. }
\end{figure}

\textbf{Phase relations near the melting temperature.}
We recently showed that the \emph{ab initio} melting temperature could be accurately obtained from the AIMD simulation with the help of a semi-empirical potential if the potential provides atomic structures close to those obtained in the AIMD simulation \cite{10}. Therefore, we developed a Finnis-Sinclair (FS) type potential \cite{39} for Ni in the present study. The Supplementary Information contains details of potential and accuracy tests. The bcc, hcp, and fcc melting temperatures for this FS potential are computed by two-phase coexistence simulations with large-scale classical MD. While no information about the melting temperatures was included in the potential development, the FS potential predicts that bcc-Ni has the highest melting temperature. We calculated the \emph{ab initio} free energy via thermodynamic integration (TI), using the classical system described by the FS potential as the reference state (see Method). Table~\ref{table:tab1} shows the \emph{ab initio} melting temperatures of the three phases computed at 323 GPa and 360 GPa, which are the pressures at the boundary and center of the inner core, respectively. Bcc-Ni indeed shows higher melting temperatures than hcp-Ni and fcc-Ni. The differences are small, $\sim$ 30 K, but are more significant than the typical confidence interval of 15 K in similar calculations \cite{10}. The difference between hcp-Ni and fcc-Ni melting temperatures is smaller, $\sim$7 K. In Table~\ref{table:tab1} we compare the melting temperatures of Ni phases to those of Fe phases computed in Ref. \cite{10}. Both bcc Ni and hcp Ni's melting temperatures are significantly higher than those for Fe phases by 700-800K.

\begin{table}[t]
\caption{\label{table:tab1} Melting temperatures of Fe and Ni phases at 323 GPa and 360 GPa obtained from \emph{Ab initio} calculations. Fe data are taken from Ref. \cite{10}. The uncertainty is 15 K.}
\centering
\begin{tabular} {c | c | c | c | c}
\hline
\hline
System       & P (GPa) &  $T_m^{bcc} (K)$ &  $T_m^{hcp} (K)$ & $T_m^{fcc} (K)$    \\
\hline
Ni 	& 323 & 6528	& 6499 &	6492 \\
\hline
Ni 	& 360 & 6870	& 6833	& 6824 \\
\hline
Fe 	& 323 & 5632	& 5848	& - \\
\hline
Fe 	& 360	& 5850	& 6094	&- \\
\hline
\hline
\end{tabular}
\end{table}

Figure~\ref{fig:fig2}a shows the relative free energy difference referenced to the bcc phase at 323 GPa. Bcc is the stable phase from 6528 K to 6366 K. For temperatures lower than 6366 K, hcp is the stable phase while bcc becomes metastable. A comparison between Fig.~\ref{fig:fig2}a and~\ref{fig:fig2}b shows that the bcc-hcp phase relations are very different between Fe and Ni. The bcc phase is always metastable for Fe, while it is stable for Ni in a small temperature range near the melting point. The free energy difference between hcp and bcc is much smaller for Ni than for Fe. Therefore, Ni is likely a bcc stabilizer if mixed with Fe at high temperatures under core pressures.

\begin{figure}
\includegraphics[width=0.49\textwidth]{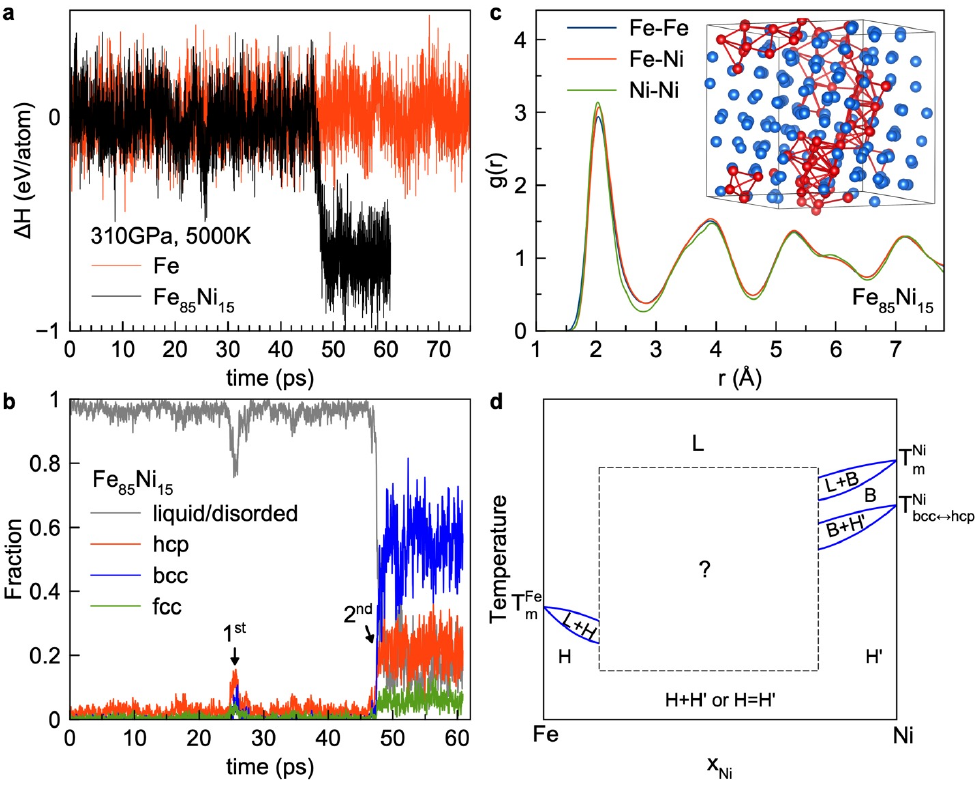}
\caption{\label{fig:fig3} \textbf{Fe and Fe$_{85}$Ni$_{15}$ liquid at 310 GPa and 5000 K by the AIMD simulation.} a. The enthalpy change as a function of simulation time. The data are referenced to averaged liquid enthalpy for Fe and Fe$_{85}$Ni$_{15}$, respectively. b. The fraction of liquid/disordered and crystalline atoms as a function of time in Fe$_{85}$Ni$_{15}$ simulation. The arrows mark the two nucleation attempts. c. The partial pair correlation function of the crystallized Fe$_{85}$Ni$_{15}$ phase averaged over the last 10 ps of the simulation. The insert shows the final atomic configuration. The center atoms of hcp-like clusters are colored red and connected to guide the eyes. Blue are bcc. d. Schematic of a “likely” Fe-Ni liquid-bcc-hcp phase diagram. Ni ($T_m^{Ni}$) has a higher melting temperature than Fe ($T_m^{Fe}$). For pure Fe, hcp is the stable phase, and bcc is metastable. For pure Ni, bcc is the high-tempeature stable phase and transform to hcp at lower temperatures. L stands for liquid phase; B is bcc phase and H and H’ are hcp phases. Dashed box indicates uncertain shape/existence.}
\end{figure}

\textbf{Crystallization of Fe-Ni mixture.}
The comparisons between Fe and Ni suggest the effect of Ni on the core's structure and formation should be carefully considered. Previous studies have found liquid Fe at core pressures requires an unrealistically large undercooling to nucleate the solid inner core phase, which is unlikely to be reached under Earth's core conditions and age \cite{18,20,21}. Since Ni's melting temperature is considerably higher than Fe's at the same pressures, the Fe-Ni mixture could have an increased liquidus temperature compared to Fe's melting temperature. Because Fe and Ni liquids nucleate the bcc phase first \cite{21}, we should consider a simple metastable bcc-liquid phase diagram schematically shown in Fig.~\ref{fig:fig3}a. It demonstrates that mixing Ni in Fe can effectively change the supercooling ($\Delta T$) and might accelerate the nucleation process. We performed AIMD simulations of Fe and Fe$_{85}$Ni$_{15}$ liquids at large supercooling temperatures under core pressures to examine this mechanism. At $\sim$310 GPa and 5000 K, the Fe$_{85}$Ni$_{15}$ liquid crystallizes within 50 ps, as shown in Fig.~\ref{fig:fig3}b. In contrast, no crystallization was observed for the Fe liquid up to 76 ps under the same P-T condition. Crystallization is a stochastic process; consequently, the nucleation incubation time can fluctuate. We performed three additional AIMD simulations for Fe$_{85}$Ni$_{15}$ and Fe to counteract this stochastic effect. The results are presented in Supplementary Figure S3. We found the averaged crystallization time for Fe$_{85}$Ni$_{15}$ is $\sim$ 40 ps, while no crystallization can be observed with pure Fe up to 80 ps at 5000 K and 310 GPa. This suggests that Fe$_{85}$Ni$_{15}$ alloy has a higher nucleation rate than pure Fe. The Ni's effect on accelerating Fe's crystallization process can also be observed from large-scale classical MD simulations by cooling the melts with ultrahigh cooling rates (Supplementary Note 2). It shows that 15$\%$ Ni reduces the required supercooling by $\sim$400 K under a cooling rate of $10^{11}$ K/s. While this cooling rate is far from Earth's core condition, it validates Ni's effect on Fe's crystallization revealed by the AIMD simulations. We note that Fig.~\ref{fig:fig3}a may be changed with the inclusion of light elements, while the effect of Ni should always be considered to study the nucleation of the core.

The crystallization process of Fe$_{85}$Ni$_{15}$ liquid in the AIMD simulations is analyzed based on its local atomic structure change in Fig.~\ref{fig:fig3}c. Interestingly, the liquid shows two attempts to nucleate: first at $\sim$25 ps, then at 47 ps. In the first attempt, an hcp nucleus emerges but only grows to a small size and then remelts. This indicates that the nucleus cannot overcome the nucleation barrier to reach the critical size. In the second attempt, a bcc nucleus emerges and successfully crystallizes. During the bcc growth, $\sim$20$\%$ of liquid atoms transform to hcp. It forms a bcc-hcp coexisting system in the as-crystallized solid, shown in Fig.~\ref{fig:fig3}d. The partial pair correlation functions of Fe and Ni are almost indistinguishable, suggesting Fe and Ni atoms randomly distribute in the as-crystallized solid. This is consistent with previous calculations that Fe and Ni do not form stoichiometric compounds but only form completely disordered solid solutions under core conditions \cite{40,41}. Additional three AIMD simulations of Fe$_{85}$Ni$_{15}$ crystallization in Supplementary Figure S3 also show  minor hcp phases in the as-crystallized bcc phases.

\textbf{Discussion.}
The current results demonstrate that Ni strongly stabilizes the bcc phase and accelerates the crystallization of Fe under Earth's core conditions. Moreover, our simulation showed that bcc and hcp can coexist during the crystallization of Fe$_{85}$Ni$_{15}$ alloy. These results suggest a potential coexistence of hcp and bcc phases under Earth's inner core. Using this study's minimal free energy data, one can propose a binary Fe-Ni phase diagram, as illustrated in Fig.~\ref{fig:fig3}d. On the Ni side, the bcc phase has a stability field close to the melting curve. This bcc stability field can extend into the Fe-rich domain, posing a challenge to the stability of hcp, potentially leading to the emergence of a liquid-hcp-bcc eutectic. Thus, there's a possibility for a thermodynamically stable coexistence of bcc and hcp in the solid phase of the inner core. At lower temperatures, when both Fe and Ni are in the hcp phase, they might form two distinct hcp phases (H+H') with varying Ni compositions or a single one, depending on how different or similar their partial molar volumes are. The nature of the equilibrium states in Fig.~\ref{fig:fig3}d should be investigated via simulations and experiments in future studies. Interestingly, a recent experimental study revealed the coexistence of hcp and B2 phases in the Fe$_{93}$Ni$_{7}$ alloy at 186 GPa and 2970 K \cite{42}. A comprehensive Fe-Ni phase diagram encompassing liquidus, solidus, and solvus curves is essential to advance understanding of the Earth's inner core structure and the associated seismic velocity anomalies \cite{51,52}.

To conclude, we find that the bcc phase crystallizes from Ni liquid at a temperature above the melting point of Fe under inner core pressures. \emph{Ab initio} free energy calculations indicate that Ni's melting temperature is ~700-800 K higher than Fe’s at 323-360 GPa. Bcc is the thermodynamically stable phase near Ni's melting point at inner core pressures. Ni accelerates Fe's crystallization process at inner core conditions. The Fe-Ni mixture may lead to the coexistence of hcp and bcc phases under core conditions. These results suggest that Ni can be a key factor in modeling the Earth's inner core formation and present structure. 

\section{Methods}
\textbf{\emph{Ab initio} molecular dynamics simulations.} \emph{Ab initio} molecular dynamics (AIMD) were employed with the Born Oppenheimer approach to simulate crystallization, collect input data to develop semi-empirical potential and perform thermodynamic integration calculations. The Vienna \emph{ab initio} simulation package (VASP) \cite{43} was employed for the density-functional theory (DFT) calculations. The projected augmented-wave (PAW) potentials shipped with VASP was used to describe the electron-ion interaction with 8 and 10 valence electrons for Fe and Ni, respectively. The core radii cutoff of these PAW potential are 1.2 Å. The effect of the number of valence electrons was examined in the calculation of the Fe's melting temperature of the bcc and hcp phases 10. At 323 GPa, Fe hcp and bcc's melting temperatures are 5848 K and 5632 K, respectively, with PAW8 potential. They are 6357 K and 6168 K with PAW16 potential, which includes additional $3s^23p^6$ electrons. The difference between hcp and bcc T$_m$ was 216 K and 189 K for PAW8 and PAW16 potentials, respectively. Therefore, including more inner shell electrons as valence electrons can systematically increase the melting temperature \cite{10,36}. However, it does not significantly affect the relative melting temperature differences between bcc and hcp phases \cite{10}. To achieve sufficiently long simulations of crystallization, we employ the PAW potentials without  $3s^23p^6$  for both Fe and Ni, which provides a consistent condition for describing their atomic interactions. The generalized gradient approximation (GGA) in the Perdew-Burke-Ernzerhof (PBE)  form was employed for the exchange-correlation energy functional. A plane-wave basis set was used with a kinetic energy cutoff of 400 eV. The AIMD simulations were performed for the constant number of atoms, volume, and temperature (NVT) canonical ensemble. In the crystallization simulations, Ni and Fe$_{85}$Ni$_{15}$ liquids are modeled by 250 Ni atoms and Fe$_{212}$Ni$_{38}$ atoms, respectively. The $\Gamma$ point was used to sample the Brillouin zone of these supercells for crystallization. A time step of 1.5 $fs$ was used to integrate Newton's equations of motion. The Nos\'e-Hoover thermostat was employed to control the temperature. The electronic entropy at high temperatures was described by the Mermin functional \cite{44,45}. The non-magnetic calculations were performed in AIMD. We tested the spin-polarized calculations and found that the magnetic moments of Fe and Ni phases are quenched when the electronic temperature is more than 4000 K at 323 GPa. 

\textbf{Local structure characterization.} The cluster alignment (CA) method \cite{46,47} was employed to recognize bcc, fcc, and hcp short-range orders in the local atomic clusters during crystallization. The CA method aligns the atomic clusters to the standard bcc, fcc, and hcp templates and computes the root mean square deviation (RMSD) between the atomic clusters and templates. Supplementary Figure S4 indicates that CA can distinguish well crystalline phases from the liquid. To remove the noise caused by thermal fluctuations, the atomic positions were averaged for 0.06 ps to perform the CA analysis.

\textbf{Classical molecular dynamics simulations.} Classical molecular dynamics (CMD) simulations were performed with LAMMPS (Large-scale Atomic/Molecular Massively Parallel Simulator) code \cite{48}. The interatomic interaction was modeled using the Finnis-Sinclair (FS) type \cite{39} semi-empirical potential developed in this work. During the MD simulation, the constant number of atoms, pressure, and temperature (NPT) ensemble was applied with the Nosé-Hoover thermostat and barostat. The time step of the simulation was 1.0 $fs$. The melting temperatures of the classical system, denoted as $T_C^m$, were determined using the solid-liquid coexistence approach \cite{49} with 22,500 atoms. The free energy difference between liquid and solid, $\Delta G_C^{L-S}$, is determined by the Gibbs-Helmholtz equation,

\begin{equation}
\Delta G_C^{L-S} (T)=-T\int_{T_C^m}^T \Delta H(T)/T^2  dT, \label{1}
\end{equation}

where $\Delta$H is the latent heat which is computed by the enthalpy difference between liquid and solid from classical simulations with 5,000 atoms.

\textbf{\emph{Ab initio} melting temperature.} The \emph{ab initio} melting temperature, $T_A^m$, was obtained when the \emph{ab initio} free energy difference between liquid and solid phases is zero, i.e., $\Delta G_A^{L-S} (T_A^m)=0$. $\Delta G_A^{L-S}$ is computed by thermodynamic integration (TI) between the classical system (C) and \emph{ab initio} system (A) \cite{10}. It was performed by exchanging the \emph{ab initio} and classical atomic information on-the-fly in an MD simulation \cite{10}. Classical molecular dynamics simulations were performed with LAMMPS (Large-scale Atomic/Molecular Massively Parallel Simulator) code \cite{48}. During the TI-MD, the NVT ensemble was applied, and the Nos\'e-Hoover thermostat \cite{50} was employed to control the temperature. A time step of 2.0 $fs$ was used to integrate Newton's equations of motion. Supercells with 288, 256, 250, and 250 atoms were used to simulate hcp, fcc, bcc, and liquid, respectively.

\section{Acknowledgments}
Work at Iowa State University and Columbia University was supported by the National Science Foundation awards EAR-1918134 and EAR-1918126. We acknowledge the computer resources from the Extreme Science and Engineering Discovery Environment (XSEDE), which is supported by the National Science Foundation grant number ACI-1548562. X.L. and B.D. are supported by JSPS KAKENHI Grant Number JP21K14656. Molecular dynamics simulations were supported by the Numerical Materials Simulator supercomputer at the National Institute for Materials Science (NIMS). S. Fang and T. Wu from Information and Network Center of
Xiamen University are acknowledged for the help with the GPU computing.

\bibliographystyle{apsrev4-1}

\end{document}